\documentclass[journal]{IEEEtran}
\usepackage{cite}

\ifCLASSINFOpdf
\else
\fi
\usepackage{amsmath}
\usepackage{amsthm}
\usepackage{algorithm}
\usepackage{algorithmic}
\usepackage{mathrsfs}
\usepackage{amsfonts,amssymb}
\usepackage{booktabs}

\usepackage{subfigure}
\ifCLASSINFOpdf
   \usepackage[pdftex]{graphicx}
   \graphicspath{{../pdf/}{../jpeg/}}
   \DeclareGraphicsExtensions{.pdf,.jpeg,.png}
\else
   \usepackage[dvips]{graphicx}

   \graphicspath{{../eps/}}

   \DeclareGraphicsExtensions{.eps}
\fi
\ifCLASSOPTIONcompsoc
\else
  \usepackage[caption=false,font=footnotesize]{subfig}
\fi

\begin{document}
%
\title{Energy Optimization for Cellular-Connected Multi-UAV Mobile Edge Computing Systems  with Multi-Access Schemes}

%


\author{Meng~Hua,~\IEEEmembership{Student Member,~IEEE,}
        Yi~Wang,
Qingqing~ Wu,~\IEEEmembership{Member,~IEEE,}
        Chunguo~Li,~\IEEEmembership{Senior Member,~IEEE,}
        Yongming~Huang,~\IEEEmembership{Senior Member,~IEEE,}
        and~Luxi~Yang,~\IEEEmembership{Member,~IEEE,}

}

\maketitle

\begin{abstract}
In this paper, a cellular-connected  unmanned aerial vehicles (UAV) mobile edge computing system is studied in which the multiple UAVs are served by terrestrial base station (TBS) for computation task offloading. Our goal is to minimize the total UAVs energy consumption, including propulsion  energy, computation energy and communication energy,  while ensuring that the total  number of bits of  UAVs are  completely computed. For tackling the large amount of bits for computation, we  propose a resource partitioning strategy where one portion of tasks  is migrated to TBS for computation and the other portion of tasks is locally computed at UAV. For deeply comprehending the impacts of access manners on the system performance,   we consider four access schemes in the uplink transmission, i.e., time division multiple access (TDMA), orthogonal frequency division multiple access (OFDMA), One-by-One access and non-orthogonal multiple access (NOMA). The problem of jointly optimizing bit allocation, power allocation, resource partitioning as well as UAV trajectory under  TBS's energy budget is  formulated and tackled by means of successive convex approximation (SCA) technique. The numerical results show that the proposed schemes save  much energy compared with benchmark schemes.
\end{abstract}

\begin{IEEEkeywords}
Mobile edge computing, UAV trajectory, bit allocation, power allocation, resource partitioning, energy minimization.
\end{IEEEkeywords}

%
\IEEEpeerreviewmaketitle

\section{Introduction}
%
%
%
%
Recently, the use of unmanned aerial aircraft (UAV) as mobile vehicle is emerging as an effective complementary technique for future fifth generation due to the UAV's flexible mobility \cite{Zeng2016Wireless,mozaffari2018tutorial,Xiao2016Enabling,wu2018common,wu2018capacity,zhang2018securing}. In particular, the  UAV can play a key role in enabling wireless connectivity in various scenarios, e.g.,  UAV-aided  ubiquitous coverage, UAV-aided relaying and  UAV-aided information dissemination, etc \cite{alzenad20173,Mozaffari2016Efficient ,bor2016efficient,zeng2016throughput,zhao2004message,zeng2018trajectory,zeng2018energy,Wu2017Joint,yang2018energy}. In some hot spots or in some harsh areas where the base stations are difficult to be deployed, by  means of using multi-UAV  deployed as aerial  base stations (ABSs)  has become an promising solution to provide seamless wireless coverage within the serving area \cite{alzenad20173,Mozaffari2016Efficient,bor2016efficient}. The UAV used as mobile relaying provides new opportunities for system enhancement compared with conventional static relaying, since it can  successively  adjust its location for experiencing good channel condition \cite{zeng2016throughput},\cite{zhao2004message}. In addition, the UAV  can be used for dispatching  to disseminate data to the sensor nodes in internet of thing (IoT) scenario \cite{zeng2018trajectory,zeng2018energy,Wu2017Joint,yang2018energy}.

As briefly  reviewed above, the prior works about  UAV  assume that they use UAV as  either mobile relaying or ABS. However, the UAV used as \emph{flying cloudlet} in which the small cloudlet has  an abundance of computation resource as well as communication resource  is  obtaining  great attention \cite{jeong2017mobile,Jeong2018Mobile,cao2018mobile,zhou2018uav,Cheng2018UAV}.  In the traditional cellular network, the processor unit always resides in the cloud and the  distance between mobile user and cloud is always long,  which may cause long latency and power consumption for transmission in the uplink when  the mobile user located  in cellular edge. In order to reduce the latency and power consumption, some researchers have been paying attention to dealing with  it. However, by far, only a few works have studied the UAV-enabled mobile edge computing system. The authors in \cite{jeong2017mobile} proposed a  UAV-mounted cloudlet scenario where the UAV provided offloading opportunities to a single mobile user.  The goal of \cite{jeong2017mobile} was to minimize the mobile energy consumption by optimizing bit allocation with a pre-determined UAV trajectory. The same authors extended their work  \cite{jeong2017mobile}  into a more general case where the UAV was employed as cloudlet to offer computation offloading opportunities to multiple mobile users \cite{Jeong2018Mobile}. The  \cite{Jeong2018Mobile} aimed  at  minimizing the total of mobile users energy consumption by jointly optimizing bit allocation and UAV trajectory. For maintaining  the sustainable offloading task, a UAV-enabled wireless powered mobile edge computing  system was studied in \cite{zhou2018uav}. Specifically,  \cite{zhou2018uav} considered a scenario where the energy transmitter and cloudlet are  mounted on UAV, the UAV firstly transmitted energy to multiple mobile users and then  mobile users  exploited the harvested energy for computation tasks offloading. In order to extend the coverage in the cellular networks, a  cellular-connected UAV communication has being proposed  as a promising  solution where the  UAV is  integrated into cellular networks as a new  mobile user\cite{mozaffari2018tutorial}, \cite{cao2018mobile}, \cite{lyu2017blocking}, \cite{zhang2017cellular}. In addition, the emerging diverse mobile applications such as virtual reality (VR), high definition videos, mobile online gaming, etc., which  require low  latency and  high computation capability. To tackle this issue, the work in \cite{cao2018mobile} established a cellular-connected UAV networks, the goal of this work was to  minimize the UAV's mission completion time by jointly optimizing UAV  trajectory and time allocation.

However, all the above mentioned works have not addressed the energy consumption problem of  UAVs for computation tasks offloading.  On the one hand, due to the on-board battery capacity  is constrained by the limited size of   UAVs,  the energy consumption of UAVs becomes  a huge challenging in a cellular-connected UAV system. Therefore, how to prolong the UAV flying time become an open problem.  It should be pointed out that we should carefully consider three parts of energy consumption of UAV, including  communication-related energy consumption, computation energy consumption and propulsion-related energy consumption \cite{zeng2018energy},  \cite{faqir2017joint,zeng2017energy,8329973, Zhang2016Fundamental,Wu2017An,wu2018fundamental}. On the other  hand, for deeply comprehending the impacts of access manners on the system performance, the  multi-access schemes should be carefully designed. Based on this fact, in this paper, we study a  cellular-connected UAV system where multi-UAV  simultaneously communicate with one TBS for computation task offloading under different access  schemes.  Different from  previous works \cite{jeong2017mobile,Jeong2018Mobile,cao2018mobile,zhou2018uav} where  the authors assume that the computation tasks can be completely migrated from mobile users to TBS for  computing, whereas it will not work  as the number of bits for computing is very large. Indeed, the   number of bits used for offloading for computing   is constrained by many factors, such as the TBS energy budget,  UAV transmit power, UAV-TBS channel condition, etc. For this, we propose a resource partitioning strategy where one portion of tasks  is migrated to TBS for computation in the uplink transmission  and the other portion of tasks is locally computed at UAV.
Based on these, our goal is to minimize the total UAVs energy consumption while ensuring that the total number of bits of  UAVs are completely computed in a given horizon time. Since the formulated problem is in a non-convex form, which cannot be efficiently solved by the standard optimization technique. To tackle this non-convex problem, we  obtain a sub-optimal solution by leveraging successive convex approximation (SCA) technique. To the best  of our knowledge, it is the first work to investigate the optimization of offloading process from  multi-UAV to TBS  by jointly considering  the communication-related energy consumption, local computation energy consumption and propulsion-related energy consumption. The main contributions of this paper can be summarized as follows:
\begin{itemize}
\item We study  a model for multi-UAV mobile edge computing system,  where the tasks at UAVs are migrated to TBS for computation. For deeply comprehending the impacts of access manners on the system performance, we present four access schemes in the uplink transmission in our model, i.e., time division multiple access (TDMA), orthogonal frequency division multiple access (OFDMA), One-by-One access and non-orthogonal multiple access (NOMA). An energy minimization of UAVs is then formulated  to jointly optimize bit allocation, UAV trajectory, UAV power allocation and resource partitioning with a given horizon time, subject to practical UAV mobility, transmit power and \emph{bit-casuality } constraints.
\item  For tackling the large number of bits for computation,  we propose a resource partitioning strategy where one portion of tasks  is migrated to TBS for computation in the uplink transmission  and the other portion of tasks is locally computed at UAV, which has been verified the effectiveness of this strategy  in Section VII.
\item Since the formulated problems are in different forms under different access schemes,  we solve these formulated problems  separately for different  access schemes. For TDMA and OFDMA schemes, we obtain a sub-optimal solution by leveraging SCA technique. For One-by-One and NOMA schemes, we decompose the original problem into two subproblems, and develop an efficient iterative algorithm by optimizing  the two  subproblems alternately. The numerical results show that our proposed four schemes save a large amount of energy compared with propulsion minimization scheme. What's more, while the NOMA scheme saves more energy consumption of UAVs compared with TDMA, OFDMA and One-by-One scheme, the successive interference cancellation (SIC) technique used for NOMA, which may bring an additional interference and implementation complexity of NOMA.
\end{itemize}
 \begin{figure}[!t]
\centerline{\includegraphics[width=3in]{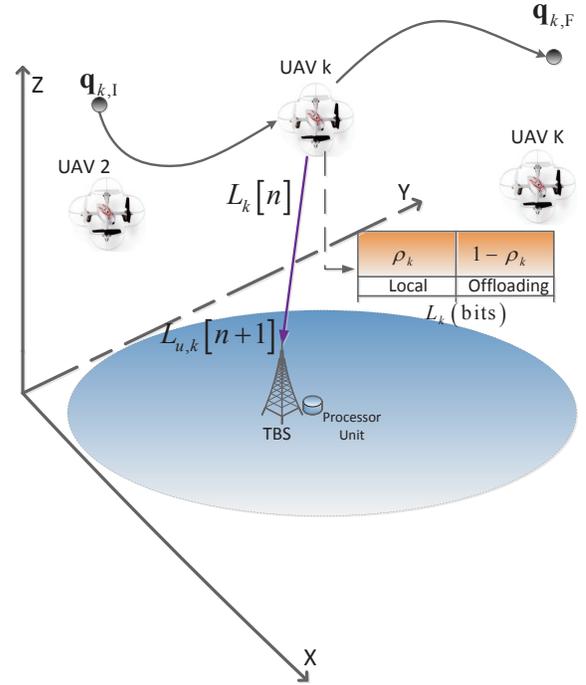}}
\caption{Cellular-Connected Multi-UAV Mobile Edge Computing Systems.} \label{fig1}
\end{figure}
The rest of this paper is organized as follows. Section II describes the system model and introduces the energy consumption of UAV  in mathematical expression. In Section III, we formulate the UAV energy consumption  minimization problem based on TDMA scheme, and obtain a sub-optimal solution by applying SCA technique. In Section IV, we formulate the UAV energy consumption  minimization problem based on OFDMA scheme, and obtain a sub-optimal solution by  applying SCA technique. In Section V, we formulate the UAV energy consumption  minimization problem based on One-by-One scheme, we decompose the original problem into two subproblems, and develop an efficient iterative algorithm by optimizing  the UAV trajectory and UAV-TBS scheduling  alternately. In Section VI, we formulate the UAV energy consumption  minimization problem based on NOMA scheme, we decompose the original problem into two subproblems, and develop an efficient iterative algorithm by optimizing  the UAV trajectory and UAV transmission power   alternately. In Section VII, the simulation results are given to validate the performance of our proposed schemes. Finally, we conclude the paper in Section VIII.

%
%

\section{SYSTEM MODEL}
As shown in Fig. 1, we consider an uplink transmission scenario where a set $\cal K$ of $K$ single-antenna UAVs are serviced by one single-antenna TBS for tasks offloading. We study the optimization of offloading process from the UAVs to TBS with the goal of minimizing the total energy consumption of all UAVs for a finite time horizon $\rm T$. In Fig. 1, we can see that one portion of  total bits of UAV $k$, $\rho_k$, is computed locally  at UAV $k$, and the other portion of  bits, $1-\rho_k$, is migrated to TBS for computation. In particular, $\rho_k=0$ means that  all the  number of bits at UAV $k$ are  completely migrated to TBS for computation, and  $\rho_k=1$ indicates  that  all the  number of bits at UAV $k$ are  used for locally computing.

For the sake of mathematical description simplicity, we consider a three-dimensional Cartesian coordinate system, with all dimensions being measured in meters, where  the horizontal coordinate of UAV $k$ at time instant $t$ is denoted by  ${{\bf{q}}_k}\left( t \right) = {\left( {{x_k}\left( t \right),{y_k}\left( t \right)} \right)^T} \in {\mathbb R^{2 \times 1}}$. Without loss of generality, the TBS is located at origin of the horizonal coordinate, denoted as  ${\bf w}$. We assume that the UAVs move at a fixed altitude $H$ under the initial/final location, maximization UAV speed and acceleration constraints. For tackling the problem more tractable, the continuous horizon time $\rm T$ is equally divided into  $N+1$ time slots with each slot  duration $\delta $. As such,  the UAV trajectory ${\bf{q}}_k\left( t \right)$, UAV speed ${\bf{v}}_k\left( t \right)$, UAV acceleration ${\bf{a}}_k\left( t \right)$   over time $\rm T$  can be approximately denoted by  $N \text {+}2$-length sequences as $\left\{ {{\bf q}_k\left[ n \right]} \right\}_{n = 0}^{N+1}$, $\left\{ {{\bf v}_k\left[ n \right]} \right\}_{n = 0}^{N+1}$, $\left\{ {{\bf a}_k\left[ n \right]} \right\}_{n = 0}^{N+1}$, where ${\bf q}_k\left[ n \right] = {\bf q}_k\left( {n\delta } \right)$, ${\bf v}_k\left[ n \right] = {\bf v}\left( {n\delta } \right)$, ${\bf a}_k\left[ n \right] = {\bf a}_k\left( {n\delta } \right)$, $k \in \cal K$.

It can be  assumed that the  channels between the UAVs and TBS are dominated by line-of-sight (LoS) without small-scale  fading, which have been verified  by taking measurements in \cite{matolak2017air} and \cite{lin2017sky}. Thus, the channel power gain from UAV $k$ to TBS in the uplink at time slot $n$ can be  modeled as \cite{zeng2016throughput},\cite{ Wu2017Joint},\cite{zeng2017energy},
\begin{equation}
{h_k}\left[ n \right] = \frac{{{\beta _0}}}{{d_k^2\left[ n \right]}} = \frac{{{\beta _0}}}{{{{\left\| {{\bf{q}}_k\left[ n \right] - {{\bf{w}}}} \right\|}^2} + {H^2}}},
\end{equation}
where ${d_k\left[ n \right]}$ denotes the distance between UAV $k$ and TBS at time slot $n$, $\beta_0$ represents the reference channel gain at $d=1 \rm{m}$.

Next, we will  discuss the energy consumption models of UAV. As it is mentioned before, the total energy consumption of UAV consists of three  parts, i.e., UAV's computation  energy, UAV's communication energy and UAV's propulsion energy, which are described in below.
\subsubsection{Execution Energy Consumption  Model}
The energy consumption of local computation is directly determined by the CPU workload $W$, which is related to the data size and the complexity algorithm of application \cite{Miettinen2010Energy}, \cite{zhang2013energy}, and is defined as
\begin{align}
W=LC,
\end{align}
where $L$ denotes the input  data size, and $C$ represents the complexity of the application, which has been shown that follows  Gamma distribution \cite{zhang2013energy}, \cite{lorch2001improving}. Three factors can be  contributed to the CPU power consumption including dynamic power, short circuit power and leakage power. As stated in \cite{yuan2006energy}, the energy consumption of CPU is dominated by dynamic power and other two factors can be neglected. In order to minimize the energy consumption of CPU power (dynamic power), the optimal clock-frequency scheduling in each CPU cycle is  achieved in \cite{zhang2013energy},\cite{sheng2015energy}. Let denote the number of bits  of UAV $k$ as $L_k$, $k \in \cal K$. Based on \cite{zhang2013energy},\cite{sheng2015energy},  the  energy computation consumption model with data size $\rho_k L_k$ and completion execution  time $\rm T$ can be expressed as
\begin{align}
E_k^{comp} = \frac{{G{{(\rho_k L_k)}^3}}}{{ {\rm{T}^2}}}, \label{computenergy}
\end{align}
where the coefficient $G$ is a constant that  accounts for   the  effective switched capacity and application execution completion probability.
As we can see from formula \eqref{computenergy}, the increased number of bits or the reduced  time ${ {\rm{T}}}$ will drastically increase the  computation energy consumption.
\subsubsection{Communication Energy Consumption Model for Offloading}
The required transmission  power $p_k[n]$ of UAV $k$ for  completely transmitting $L_k[n]$ bits in the uplink within time slot $n$ should satisfy  Shannon theory, we thus have
\begin{align}
B\delta {\log _2}\left( {1 + \frac{{{p_k}[n]{h_k}\left[ n \right]}}{{{\sigma ^2}}}} \right) = L_k[n], \label{Shannon}
\end{align}
where $B$  denotes the system bandwidth and ${\sigma ^2}$ is the additive white Gaussian noise power.
After some algebraic manipulations,  the transmit power $p_k[n]$ can be expressed as
\begin{align}
{p_k}\left[ n \right] = \left( {{2^{\frac{L_k[n]}{{B\delta }}}} - 1} \right)\frac{{\sigma ^2}}{{{h_k}\left[ n \right]}}.
\end{align}
As such, the required energy of UAV $k$ for  completely transmitting $L_k[n]$ bits within time slot $n$  can be obtained as
\begin{align}
E_k^{com}[n]=p_k[n]\delta, \label{comenergy}
\end{align}
\subsubsection{Propulsion-related Energy Consumption of UAV  for Flying}
Based on the model in \cite{Jeong2018Mobile}, \cite{zeng2017energy}, the total  UAV's propulsion energy consumption of UAV over $\rm T$  can be expressed as
\begin{equation}
{E_{k}^{fly}} = \sum\limits_{n = 1}^N {\left( {{c_1}{{\left\| {\textbf{v}\left[ n \right]} \right\|}^3} + \frac{{{c_2}}}{{\left\| {\textbf{v}\left[ n \right]} \right\|}}\left( {1 + \frac{{{{\left\| {\textbf{a}\left[ n \right]} \right\|}^2}}}{{{g^2}}}} \right)} \right)\delta }  + {\triangle\rm{p}}, \label{flyenergy}
\end{equation}
where  ${\triangle\rm{p = }}\frac{1}{2}m\left( {{{\left\| {\textbf{v}\left( T \right)} \right\|}^2} - {{\left\| {\textbf{v}\left( 0 \right)} \right\|}^2}} \right)$ represents the change of UAV's kinetic energy, which is a invariant scalar with fixed final/initial location of UAV, and $c1>0$ and  $c2>0$  are constants which are related to the UAV' wing area, load factor and wing span efficiency etc. Without loss of generality, we assume that  the UAV's initial speed and final speed are  same, thus, ${\triangle\rm{p  }}$ can be omitted in term \eqref{flyenergy}. In addition,  the formula \eqref{flyenergy} shows that the UAV's  propulsion energy is related to  UAV velocity and acceleration. Especially, as UAV hovers over a fixed location, i.e., $\left\| {\textbf{v}\left[ n \right]} \right\| = 0$, then  ${E_k^{fly}}  $ tends to be infinite.

In this paper, four access frame structures are discussed, namely TDMA, OFDMA, One-by-One access and NOMA. For the frame structure of TDMA scheme, each time slot duration $\delta$ is equally divided into $K$ sub-slot with duration $\delta_1$, i.e., ${\delta_1} =\delta/K$. However, each UAV shares the same bandwidth $B$ with each sub-slot; For the frame structure of OFDMA scheme, each UAV occupies the same time slot duration, whereas  shares the bandwidth $B/K$; For the frame structure of One-by-One scheme, we assume that at most one UAV can be scheduled within each time slot; For the  frame structure of NOMA scheme, each UAV shares the same time and same bandwidth.

In addition, it should be mentioned that the \emph{bit-casuality} in the processed phase at TBS must be carefully designed, i.e., the TBS can only process the bits that have already been received from UAVs \cite{Jeong2018Mobile}, \cite{zhou2018uav}. Without loss of generality, we assume that the processing delay at TBS is one slot. Furthermore, due to the fact that the size of computation outcome data in general is much smaller than that of the computation input data. Thus, the downlink transmission for bit allocation  is  ignored in this paper \cite{cao2018mobile}, \cite{Wang2017Computation}, \cite{Chen2015Decentralized}.
\section{ Energy minimization for TDMA Scheme}
In this section, we tackle the problem of minimizing the total energy consumption of UAVs for offloading by assuming the system operated in TDMA scheme. Specifically, the joint  bit allocation, resource partitioning, power allocation and UAV trajectory under the TBS's budget energy and UAV mobility constraints are optimized for minimizing the total energy consumption of UAVs. As each time slot  $\delta$ is equally split into $K$ sub-slot with duration $\delta_1$ ($\delta_1=\delta/K$), the communication-related energy consumption of UAV $k$ within time slot $n$, denoted as $\bar E_k^{com}[n]$, can be obtained by substituting the $\delta$ in \eqref{comenergy} with $\delta_1$. For notation brevity, we define bit allocation as $A = \left\{ {{L_k}\left[ n \right],{L_{u,k}}\left[ {n + 1} \right]} \right\}_{n = 1}^{N - 1}$, UAV trajectory as $B = \left\{ {{{\bf{q}}_k}\left[ n \right],{{\bf{v}}_k}\left[ n \right],{{\bf{a}}_k}\left[ n \right]} \right\}_{n = 0}^{N+1}$, set ${\cal N} \in \left\{ {1, \cdots ,N - 1} \right\}$ and set ${\cal N}_1 \in \left\{ {0, \cdots ,N } \right\}$. In this case, the energy minimization problem can be reformulated as
\begin{subequations}
\begin{align}
&\text{(P0)}:\mathop {\min }\limits_{\scriptstyle\left\{ {A,B,} \right\}\hfill\atop
\scriptstyle\left\{ {{\rho _k},{p_k}\left[ n \right]} \right\}\hfill} \sum\limits_{k = 1}^K {E_k^{fly} + \sum\limits_{k = 1}^K { E_k^{comp} + \sum\limits_{k = 1}^K {\sum\limits_{n = 1}^{N - 1} {\bar E_k^{com}\left[ n \right]} } } }  \\
&{\rm{s}}{\rm{.t}}{\rm{.}}{\kern 1pt} {\kern 1pt} {\kern 1pt} {\kern 1pt} {\kern 1pt} {\kern 1pt} {\kern 1pt} {\kern 1pt} {\kern 1pt} {\kern 1pt} {\kern 1pt} {\kern 1pt} {\kern 1pt} {\kern 1pt} {\kern 1pt} {\kern 1pt} \sum\limits_{n = 1}^{N - 1} {\frac{{G{{\left( {\sum\limits_{k = 1}^K {{L_{u,k}}\left[ {n + 1} \right]} } \right)}^3}}}{{{\delta ^2}}}} {\kern 1pt} {\kern 1pt} {\kern 1pt}  \le {E_{total}}{\kern 1pt} {\kern 1pt} {\kern 1pt} {\kern 1pt},\label{C1}\\
& {\log _2}\left( {1 + \frac{{{p_k}\left[ n \right]{h_k}\left[ n \right]}}{{{\sigma ^2}}}} \right) \ge \frac{{{L_k}\left[ n \right]}}{{{\delta _1}B}},\forall k,n \in {\cal N},\label{C2}\\
&\sum\limits_{i = 1}^n {{L_{u,k}}\left[ {i + 1} \right]}  \le \sum\limits_{i = 1}^n {{L_k}} \left[ i \right],{\kern 1pt} n \in {\cal N},\label{C3}\\
&\sum\limits_{n = 1}^{N - 1} {{L_k}} \left[ n \right] = \left( {1 - {\rho _k}} \right){L_k},\forall k,\label{C4}\\
&\sum\limits_{n = 1}^{N - 1} {{L_{u,k}}} \left[ {n + 1} \right] = \left( {1 - {\rho _k}} \right){L_k},\forall k,\label{C5}\\
&{\kern 1pt} {{\bf{q}}_k}\left[ {n + 1} \right] = {{\bf{q}}_k}\left[ n \right] + {{\bf{v}}_k}\left[ n \right]\delta  + \frac{1}{2}{{\bf{a}}_k}\left[ n \right]{\delta ^2},\forall k,n \in {{\cal N}_1},\notag\\
&{{\bf{v}}_k}\left[ {n + 1} \right] = {{\bf{v}}_k}\left[ n \right] + {{\bf{a}}_k}\left[ n \right]\delta ,n \in {\cal N}_1,\notag\\
&{{\bf{q}}_k}\left[ 0 \right] = {{\bf{q}}_{k,\rm{I}}},{{\bf{q}}_k}\left[ {N + 1} \right] = {{\bf{q}}_{{k\rm{,F}}}},\notag\ \\
&\left\| {{{\bf{v}}_k}\left[ n \right]} \right\| \le {V_{\max }},\left\| {{{\bf{a}}_k}\left[ n \right]} \right\| \le {a_{\max }},n \in {{\cal N}_1},\label{C6}\\
&0 \le {\rho _k} \le 1,\forall k,\label{C7}\\
&0 \le {p_k}\left[ n \right] \le {P_{\max }},\forall k,n \in {\cal N},\label{C8}\\
&{\kern 1pt} {\kern 1pt} {L_k}\left[ n \right] \ge 0,{\kern 1pt} {\kern 1pt} {\kern 1pt} {\kern 1pt} {L_{u,k}}\left[ {n + 1} \right] \ge 0,\forall k,n \in {\cal N}\label{C9}.
\end{align}
\end{subequations}
Eq.\eqref{C1} represents the TBS energy budget  allocated to UAVs for computation. Note that in $\text{(P0)}$, the constraint \eqref{C2} is obtained from \eqref{Shannon} by replacing the equality sign with inequality constraints. Indeed, this will not change the  optimal solution to  problem $\text{(P0)}$. To see this, suppose that the optimal solution to  problem $\text{(P0)}$ is satisfied with the strict inequality, one can always decrease the power allocation $p_k[n]$ to obtain a strictly less objective value. As a result,  at the optimal solution to $\text{(P0)}$, the  constraint \eqref{C2} must be satisfied with equality. \eqref{C3}  denotes the \emph{bit-casuality} constraint. \eqref{C4} represents that the total number of bits for transmission in the uplink, and \eqref{C5} denotes that the transmitted bits  are completely processed by TBS. \eqref{C6} represents  the UAV trajectory constraints. \eqref{C7}-\eqref{C9} are the feasible and boundary constraints of the optimization variables. We can see that the constraints set \eqref{C1} and \eqref{C3}-\eqref{C9} are all convex, however, the constraint set \eqref{C2} and objective function are not convex, which cannot be solved by standard convex technique. To proceed, we first introduce slack  variable $\left\{ {{\tau _n}} \right\}$ in term $E_k^{fly}$, then the UAV's  propulsion energy consumption, denoted as   $\bar E_k^{fly}$, can be reformulated as
\begin{align}
\bar E_k^{fly}\left[ n \right] = \left( {{c_1}{{\left\| {{\bf{v}}\left[ n \right]} \right\|}^3} + \frac{{{c_2}}}{{{\tau _n}}}\left( {1 + \frac{{{{\left\| {{\bf{a}}\left[ n \right]} \right\|}^2}}}{{{g^2}}}} \right)} \right)\delta ,
\end{align}
with additional constraint
\begin{align}
&{\left\| {{\bf{v}}\left[ n \right]} \right\|^2} \ge \tau _n^2,n \in \{ {1, \cdots ,N }\},\notag\\
&{\tau _n} \ge 0,n \in \{ {1, \cdots ,N }\}. \label{slackspeed}
\end{align}
However, the new  first constraint of  \eqref{slackspeed} is  not convex.  To tackle this non-convex constraint, we develop an SCA technique. Specifically,  with a given local point $\left\{ {{\textbf{v}_l}\left[ n \right]} \right\}$ over $l \text{-} \rm {th}$ iteration, we have
\begin{align}
{\left\| {\textbf{v}\left[ n \right]} \right\|^2}& \ge {\left\| {{\textbf{v}_l}\left[ n \right]} \right\|^2} + 2\textbf{v}_l^{\rm T}\left[ n \right]\left( {\textbf{v}\left[ n \right] - {\textbf{v}_l}\left[ n \right]} \right) \overset{a}{=}{f^{{lb}}}\left( {\textbf{v}\left[ n \right]} \right),\label{expandslackspeed}
\end{align}
where \eqref{expandslackspeed} follows from the fact that its first-order Taylor expansion of convex function is a global under-estimator \cite{bertsekas1999nonlinear}. It can be  seen that ${f^{{\rm{lb}}}}\left( {\textbf{v}\left[ n \right]} \right)$ is a linear function with respect to   $\textbf{v}\left[ n \right]$, which is convex. As such, the constraint \eqref{slackspeed} can be rewritten as
\begin{align}
&{f^{{lb}}}\left( {{\bf{v}}\left[ n \right]} \right) \ge \tau _n^2,n \in \{ {1, \cdots ,N }\},\notag\\
&{\tau _n} \ge 0,n \in \{ {1, \cdots ,N }\}. \label{NEWspeed}
\end{align}
In the next, we tackle the non-convex constraint \eqref{C2}. To this end, we first relax constraint \eqref{C2}  by introducing slack variable  $\left\{ {{y_k}\left[ n \right]} \right\}$, which can reformulate constraint \eqref{C2} as
\begin{align}
{\log _2}\left( {1 + \frac{{{p_k}\left[ n \right]{\gamma _0}}}{{{y_k}\left[ n \right] + {H^2}}}} \right) \ge \frac{{{L_k}\left[ n \right]}}{{{\delta _1}B}},\forall k,n \in {\cal N} \label{slackC2}
\end{align}
and
\begin{align}
{\left\| {{\bf{q}}_k\left[ n \right] - {w}} \right\|^2} \le {y_k}\left[ n \right],\forall k,n \in \cal N, \label{trajectory}
\end{align}
where ${\gamma _0} = \frac{{{\beta _0}}}{{\sigma ^2}}$ denotes the reference signal-to-noise (SNR). Similarly, by taking the  first-order Taylor expansion of left hand side (LHS) in constraint \eqref{slackC2} with given local point $y_{l,k}[n]$ over $l \text{-} \rm {th}$ iteration, the following global lower bound can be obtained as
\begin{align}
{\log _2}\left( {1 + \frac{{{p_k}\left[ n \right]{\gamma _0}}}{{{y_k}\left[ n \right] + {H^2}}}} \right) \ge {A_k}\left[ n \right] - r_k^{up}\left[ n \right] \overset{b}{=} R_k^{lb}\left[ n \right], \label{expandslackC2}
\end{align}
where ${A_k}\left[ n \right] = {\log _2}\left( {{y_k}\left[ n \right] + {H^2} + {p_k}\left[ n \right]{\gamma _0}} \right)$ and $r_k^{up}\left[ n \right] = {\log _2}\left( {{y_{l,k}}\left[ n \right] + {H^2}} \right) + \frac{{{{\log }_2}e}}{{{y_{l,k}}\left[ n \right] + {H^2}}}\left( {{y_k}\left[ n \right] - {y_{l,k}}\left[ n \right]} \right)$.

By replacing the non-convex objective function and \eqref{C2}  of  $\text{(P0)}$ with their corresponding lower bounds at given local point obtained above, we have the following optimization problem
\begin{align}
&\text{(P0.1)}:\mathop {\min }\limits_{\scriptstyle\left\{ {A,B,y_k[n]} \right\}\hfill\atop
\scriptstyle\left\{ {{\rho _k},{p_k}\left[ n \right]},\tau_n \right\}\hfill} \sum\limits_{k = 1}^K {\bar E_k^{fly} + \sum\limits_{k = 1}^K {E_k^{comp} + \sum\limits_{k = 1}^K {\sum\limits_{n = 1}^{N - 1} {\bar E_k^{com}\left[ n \right]} } } } \notag \\
&~~{\rm{s.t.}}~~ \eqref{C1},\eqref{C3}\text{-}\eqref{C9},\eqref{NEWspeed}, \eqref{trajectory},\eqref{expandslackC2}.\notag
\end{align}
It can be verified that the problem $\text{(P0.1)}$ is a convex optimization problem, which can be efficiently solved by standard convex technique. By successively updating the local point at each iteration via solving $\text{(P0.1)}$, an efficient algorithm is obtained for the non-convex optimization problem $\text{(P0)}$.
\section{ Energy minimization  for OFDMA Scheme}
In this section, we study the design of bit allocation, power allocation, resource partitioning and UAV trajectory for OFDMA scheme. In OFDMA system, the total bandwidth $B$ is equally divided into $K$ sub-bandwidth, each with a bandwidth of $B_0=B/K$. The formulated problem, denoted as $\text{P1}$, can be obtained by substituting the new bandwidth allocation $B=B_0$ and time allocation $\delta_1=\delta$ in  $\text{P0}$.  We summarized the resulting problem as
\begin{align}
&\text{(P1)}:\mathop {\min }\limits_{\scriptstyle\left\{ {A,B,} \right\}\hfill\atop
\scriptstyle\left\{ {{\rho _k},{p_k}\left[ n \right]} \right\}\hfill} \sum\limits_{k = 1}^K {E_k^{fly} + \sum\limits_{k = 1}^K {E_k^{comp} + \sum\limits_{k = 1}^K {\sum\limits_{n = 1}^{N - 1} {\tilde E_k^{com}\left[ n \right]} } } }  \notag \\
&{\rm{s.t.}} ~~{\log _2}\left( {1 + \frac{{{p_k}\left[ n \right]{h_k}\left[ n \right]}}{{{\sigma ^2}}}} \right) \ge \frac{{{L_k}\left[ n \right]}}{{{\delta}B_0}},\forall k,n \in {\cal N},\notag\\
&\qquad \eqref{C1}, \eqref{C3}\text{-}\eqref{C9}, \notag
\end{align}
where $ \tilde E_k^{com}[n]= \left( {{2^{\frac{L_k[n]}{{B_0\delta }}}} - 1} \right)\frac{{\sigma ^2}}{{{h_k}\left[ n \right]}} \delta$. The problem $\text{(P1)}$ is a non-convex problem, whereas can be obtain approximation solution by taking   same  manipulation operations of $\text{(P0)}$ on  $\text{(P1)}$, and the detailed procedures are omitted here for brevity.
\section{ Energy minimization for One-by-One Scheme}
In this section, we consider One-by-One access scheme in the uplink transmission.  For One-by-One access scheme, each time slot with duration $\delta$ can only be occupied by at most one UAV.  We define  a UAV-TBS scheduling indicator binary variable  ${x_k}\left[ n \right]$,  if UAV $k$ is serviced by TBS at time slot $n$, then ${x_k}\left[ n \right] = 1$, otherwise ${x_k}\left[ n \right] = 0$. It is noteworthy that within each time slot  $n$, there is at most one UAV can be scheduled, i.e., $\sum\limits_{k = 1}^K {{x_k}\left[ n \right]}  \le 1$. The corresponding design problem can be formulated as follows:
\begin{subequations}
\begin{align}
\text{(P2)}:&\mathop {\min }\limits_{\scriptstyle\left\{ {A,B,x_k[n]} \right\}\hfill\atop
\scriptstyle\left\{ {{\rho _k},{p_k}\left[ n \right]} \right\}\hfill} \sum\limits_{k = 1}^K {E_k^{fly} + \sum\limits_{k = 1}^K {E_k^{comp} + \sum\limits_{k = 1}^K {\sum\limits_{n = 1}^{N - 1} {\hat  E_k^{com}\left[ n \right]} } } }  \\
{\rm{s.t.}}~~ &x_k[n]{\log _2}\left( {1 + \frac{{{p_k}\left[ n \right]{h_k}\left[ n \right]}}{{{\sigma ^2}}}} \right) \ge \frac{{{L_k}\left[ n \right]}}{{{\delta}B}},\forall k,n \in {\cal N},\label{P2C1}\\
&\sum\limits_{k = 1}^K {{x_k}\left[ n \right]}  \le 1,n \in \cal N,\label{P2C2}\\
&{x_k}\left[ n \right] \in \left\{ {0,1} \right\},\forall k,n \in \cal N,\label{P2C3}\\
& \eqref{C1}, \eqref{C3}\text{-}\eqref{C9}, \notag
\end{align}
\end{subequations}
where ${\hat  E_k^{com}\left[ n \right]}= x_k[n]{E_k^{com}\left[ n \right]}$. Note that the term $x_k[n]$ in the first constraint of problem $\text{(P2)}$ makes sure that when UAV $k$ is  in a 'mute' condition, there have no bits transmitted in the uplink; The term $x_k[n]$ in ${\hat  E_k^{com}\left[ n \right]}$ ensures that as UAV is not scheduled by TBS, the power allocation must be zero.

One can find that problem $\text{(P2)}$ is a non-convex and mixed-integer optimization problem  due to the non-convex  objective function and integer constraints \eqref{P2C1} and \eqref{P2C3}. To tackle this non-convex and mixed-integer optimization problem,  we decompose the original problem into two sub-problems, namely UAV trajectory optimization with fixed UAV-TBS scheduling and UAV-TBS scheduling optimization with fixed UAV trajectory.
\subsection{ UAV Trajectory Optimization with Fixed UAV-TBS Scheduling}
In this subsection, we consider the first sub-problem of $\text{(P2)}$, denoted as $\text{(P2.1)}$, for optimizing UAV trajectory, bit allocation, resource partitioning and power allocation by assuming that   UAV-TBS scheduling is fixed, which can be formulated as
\begin{align}
\text{(P2.1)}:&\mathop {\min }\limits_{\scriptstyle\left\{ {A,B} \right\}\hfill\atop
\scriptstyle\left\{ {{\rho _k},{p_k}\left[ n \right]} \right\}\hfill} \sum\limits_{k = 1}^K {E_k^{fly} + \sum\limits_{k = 1}^K {E_k^{comp} + \sum\limits_{k = 1}^K {\sum\limits_{n = 1}^{N - 1} {\hat  E_k^{com}\left[ n \right]} } } } \notag \\
{\rm{s.t.}}~~ & \eqref{C1}, \eqref{C3}\text{-}\eqref{C9}, \eqref{P2C1}. \notag
\end{align}
We can see that the problem $\text{(P2.1)}$ have same structure as $\text{(P0)}$. Therefore, it can be tackled with same way as $\text{(P0)}$ and the detailed manipulations can  refer to  Section III.
\subsection{ UAV-TBS Scheduling Optimization with Fixed UAV Trajectory }
In this subsection, we consider the second sub-problem of $\text{(P2)}$, denoted as $\text{(P2.2)}$, for jointly optimizing  UAV-TBS scheduling, bit allocation, resource partitioning and power allocation by assuming that   UAV trajectory  is fixed, which can be formulated as
\begin{subequations}
\begin{align}
\text{(P2.2)}:&\mathop {\min }\limits_{\scriptstyle\left\{ {A,x_k[n]} \right\}\hfill\atop
\scriptstyle\left\{ {{\rho _k},{p_k}\left[ n \right]} \right\}\hfill} \sum\limits_{k = 1}^K {E_k^{fly} + \sum\limits_{k = 1}^K {E_k^{comp} + \sum\limits_{k = 1}^K {\sum\limits_{n = 1}^{N - 1} {\hat  E_k^{com}\left[ n \right]} } } } \notag \\
{\rm{s.t.}}~~ & \eqref{C1}, \eqref{C3}\text{-}\eqref{C9}, \eqref{P2C1},\eqref{P2C2}, \eqref{P2C3}. \notag
\end{align}
\end{subequations}
Though the problem $\text{(P2.2)}$ is in a  non-convex and mixed-integer form, we can obtain a sub-optimal solution by using relaxing method and SCA technique. Firstly, we relax the binary scheduling variable $x_k[n]$ into continuous variable, as such, the constraint \eqref{P2C3} can be reformulated as
\begin{align}
0 \le  x_k[n]  \le 1,\forall k, n \in \cal N. \label{P2.21}
\end{align}
Secondly, for non-convex term ${\hat  E_k^{com}\left[ n \right]}$ in objective function, we  tackle with it by taking first-Taylor expansion of $x_k[n]p_k[n]$ at  given local points. Specifically, for any given local points $x_{l,k}[n]$ and $p_{l,k}[n]$ over $l \text{-} \rm th$ iteration, we have the following inequality
\begin{align}
{\left( {{x_k}^2\left[ n \right] + {p_k}^2\left[ n \right]} \right)} \ge {\left( {{x_{l,k}^2}\left[ n \right] + {p_{l,k}^2}\left[ n \right]} \right)} + 2{x_{l,k}}\left[ n \right]\notag\\
\left( {{x_k}\left[ n \right] - {x_{l,k}}\left[ n \right]} \right)+ 2{p_{l,k}}\left[ n \right]\left( {{p_k}\left[ n \right] - {p_{l,k}}\left[ n \right]} \right) \overset{c}{=}z_k^{lb}[n].
\end{align}
As such, the term ${\hat  E_k^{com}\left[ n \right]}$ can be replaced by its upper bound result, which is given by
\begin{align}
\tilde E_k^{com}\left[ n \right] \le \frac{{{{\left( {{x_k}\left[ n \right] + {p_k}\left[ n \right]} \right)}^2} - z_k^{lb}\left[ n \right]}}{2}\delta \overset{d}{=}\hat E_k^{com,up}\left[ n \right]. \label{P2.22}
\end{align}
We can easily see that the expression  $\hat E_k^{com,up}\left[ n \right]$ is convex  with respect to $x_k[n]$ and $p_k[n]$. For the non-convex expression $E_k^{fly}$ with respect to ${\bf v}[n]$ and ${\bf a}[n]$, we first introduce a slack variable $\tau[n]$, and then take the first-Taylor expansion of  ${\left\| {{\bf{v}}\left[ n \right]} \right\|^2}$, the detailed procedures can refer to  Section III. At last, for non-convex constraint \eqref{P2C1}, we first introduce the slack variable $s_k[n]$, as a consequence, the constraint \eqref{P2C1} can be reformulated as
\begin{align}
&{s_k}\left[ n \right] \le {\log _2}\left( {1 + \frac{{{p_k}\left[ n \right]{h_k}\left[ n \right]}}{{{\sigma ^2}}}} \right),\forall k,n \in \cal N, \label{P2.23}
\end{align}
and
\begin{align}
&{x_k}\left[ n \right]{s_k}\left[ n \right] \ge \frac{{{L_k}\left[ n \right]}}{{\delta B}},\forall k,n \in \cal N. \label{P2.2slackC1}
\end{align}
We can see that the reformulated constraints are still non-convex due to the coupled variables $x_k[n]$ and $s_k[n]$ in \eqref{P2.2slackC1}.  However, this constraint can still approximately obtained by leveraging SCA technique, we thus have
\begin{align}
\frac{{g_k^{lb}\left[ n \right] - {{\left( {x_k^2\left[ n \right] + s_k^2\left[ n \right]} \right)}^2}}}{2} \ge \frac{{{L_k}\left[ n \right]}}{{\delta B}},\forall k,n \in \cal N, \label{P2.24}
\end{align}
where $g_k^{lb}[n]$  is given in \eqref{P2.2slack2} (see the expression on top of the next page).
\newcounter{mytempeqncnt}
\begin{figure*}
\normalsize
\setcounter{mytempeqncnt}{\value{equation}}
\begin{align}
{\left( {{x_k}\left[ n \right] + {s_k}\left[ n \right]} \right)^2} &\ge {\left( {{x_{l,k}}\left[ n \right] + {s_{l,k}}\left[ n \right]} \right)^2} + 2\left( {{x_{l,k}}\left[ n \right] + {s_{l,k}}\left[ n \right]} \right)\left( {{x_k}\left[ n \right] - {x_{l,k}}\left[ n \right]} \right)  + 2\left( {{x_{l,k}}\left[ n \right] + {s_{l,k}}\left[ n \right]} \right)\left( {{s_k}\left[ n \right] - {s_{l,k}}\left[ n \right]} \right) \notag\\
&\overset{e} {=} g_k^{lb}\left[ n \right] \label{P2.2slack2}
\end{align}
\hrulefill 
\vspace*{4pt} 
\end{figure*}

By replacing the new convex constraints \eqref{NEWspeed}, \eqref{P2.21}, \eqref{P2.22}, \eqref{P2.23} and \eqref{P2.24} of $\text{(P2.2)}$ at the $l \text{-} \rm th$ iteration obtained above, we have the following optimization problem:
\begin{subequations}
\begin{align}
\text{(P2.3)}:&\mathop {\min }\limits_{\begin{array}{*{20}{c}}
{\left\{ {A,{x_k}[n],\tau [n]} \right\}{}}\\
{\left\{ {{\rho _k},{p_k}\left[ n \right],{s_k}\left[ n \right]} \right\}{}}
\end{array}} \sum\limits_{k = 1}^K {E_k^{fly} + \sum\limits_{k = 1}^K {E_k^{comp}} }    \notag \\
&\qquad \qquad \qquad \qquad \qquad+ \sum\limits_{k = 1}^K {\sum\limits_{n = 1}^{N - 1} {\hat E_k^{com,up}\left[ n \right]} } \notag \\
{\rm{s.t.}}~~ & \eqref{C1}, \eqref{C3}\text{-}\eqref{C9}, \eqref{P2C2}, \eqref{NEWspeed}, \eqref{P2.21}, \eqref{P2.22}, \eqref{P2.23},\eqref{P2.24}.  \notag
\end{align}
\end{subequations}
It can be verified that problem $\text{(P2.3)}$ is a convex optimization problem, which can be efficiently solved by standard convex technique. To reconstruct the  binary variable, we have
\begin{align}
{x_k}\left[ n \right] = \left\{ \begin{array}{l}
1{\kern 1pt} {\kern 1pt} {\kern 1pt} {\kern 1pt} {\kern 1pt} {\kern 1pt} {\kern 1pt} {\kern 1pt} {\kern 1pt} {\rm {if}} {\kern 1pt} {\kern 1pt} {\kern 1pt} {\kern 1pt} {\kern 1pt} {x_k}\left[ n \right] \ge 0,\\
0{\kern 1pt} {\kern 1pt} {\kern 1pt} {\kern 1pt} {\kern 1pt} {\kern 1pt} {\kern 1pt} {\kern 1pt} {\rm {if}} {\kern 1pt} {\kern 1pt} {\kern 1pt} {\kern 1pt} {\kern 1pt} {x_k}\left[ n \right] < 0.
\end{array} \right.\label{reconstructedbinary}
\end{align}
It should be noted that the feasible region of  $\text{(P2.2)}$ is in general a subset of that $\text{(P2.3)}$, as a result, and the optimal value of $\text{(P2.3)}$ provides an upper bound solution to that of $\text{(P2.2)}$. Based on the solutions to its two subproblems obtained by optimizing UAV trajectory and UAV-TBS scheduling via solving $\text {(P2.1)}$ and $\text {(P2.3)}$, we propose an iterative algorithm for problem  $\text {(P2)}$, which is summarized in Algorithm 1. Note that after each iteration in Algorithm 1, the objective value of $\text {(P2)}$ is monotonically non-increasing. In addition, the objective value of $\text {(P2)}$ is lower bounded by a finite value, Algorithm 1 is thus guaranteed to converge.
\begin{algorithm}
\caption{ Joint optimization of UAV-TBS scheduling and UAV  trajectory}
\label{alg1}
\begin{algorithmic}[1]
 \STATE  \textbf{Initialize} the UAV-TBS scheduling $x_k^{m}[n]$, iterative number  $m=0$, and  a maximum iterative number $M_{max}$.
 \STATE  \textbf{Repeat}
\STATE  \qquad Solve  problem $\text {(P2.1)}$ for given  UAV-TBS scheduling \\
\qquad  $x_k^{m}[n]$, and denote the optimal UAV trajectory as \\
\qquad  $B^{m+1}$.
\STATE \qquad Solve  problem $\text {(P2.3)}$ for given  UAV trajectory \\
\qquad $B^{m+1}$, and denote the optimal UAV-TBS scheduling\\
\qquad as $x_k^{m+1}[n]$.
\STATE \qquad m=m+1.
\STATE  \textbf{Until}  a maximum iterative number has been reached or convergence.
\end{algorithmic}
\end{algorithm}
 \section{ Energy minimization for NOMA Scheme}
 In this section,   we tackle the problem of minimizing the total energy consumption of UAVs for offloading by assuming the system operated in NOMA scheme. For NOMA scheme \cite{Wu2018Spectral,Ding2014On,zhu2017optimal}, each UAV occupies the same bandwidth resource and time resource. Based on this access manner,  the problem can be formulated as
 \begin{align}
&\text{(P3)}:\mathop {\min }\limits_{\scriptstyle\left\{ {A,B,} \right\}\hfill\atop
\scriptstyle\left\{ {{\rho _k},{p_k}\left[ n \right]} \right\}\hfill} \sum\limits_{k = 1}^K {E_k^{fly} + \sum\limits_{k = 1}^K {E_k^{comp} + \sum\limits_{k = 1}^K {\sum\limits_{n = 1}^{N - 1} { E_k^{com}\left[ n \right]} } } }  \notag \\
&{\rm{s.t.}} ~{\log _2}\left( {1 + \frac{{{p_k}\left[ n \right]{h_k}\left[ n \right]}}{{{\sigma ^2} + \sum\limits_{k' \ne k} {{p_{{  k'}}}\left[ n \right]{h_{{ k'}}}\left[ n \right]} }}} \right) \ge \frac{{{L_k}\left[ n \right]}}{{\delta {B}}},\forall k,n \in \cal N,\label{P3.1}\\
&\qquad \eqref{C1}, \eqref{C3}\text{-}\eqref{C9}, \notag
\end{align}
The problem $\text{(P3)}$ is different from previous problems due to the signal interference involved in \eqref{P3.1}, which is non-convex and hard to tackle. In general, there is no standard method for solving this  non-convex  problem efficiently. In the following, we develop a two-layer iterative algorithm and solve it using SCA technique. We first split problem $\text{(P3)}$ into two subproblems, namely UAV trajectory optimization with fixed power allocation and transmission  power optimization with fixed UAV trajectory. Based on the solutions obtained, an iterative algorithm is  proposed for problem $\text{(P3)}$ via alternately optimizing the above two subproblems.
\subsection{UAV Trajectory Optimization with Fixed power Allocation}
In this section, we  consider the first subproblem of $\text{(P3)}$ for optimizing UAV trajectory with fixed transmission power allocation. The problem can be written as
  \begin{align}
&\text{(P3.1)}:\mathop {\min }\limits_{\left\{ {A,B,{\rho _k}} \right\}} \sum\limits_{k = 1}^K {E_k^{fly} + \sum\limits_{k = 1}^K {E_k^{comp}} }  \notag \\
&{\rm{s.t.}} ~{\log _2}\left( {1 + \frac{{{p_k}\left[ n \right]{h_k}\left[ n \right]}}{{{\sigma ^2} + \sum\limits_{k' \ne k} {{p_{{  k'}}}\left[ n \right]{h_{{ k'}}}\left[ n \right]} }}} \right) \ge \frac{{{L_k}\left[ n \right]}}{{\delta {B}}},\forall k,n \in \cal N, \notag\\
&\qquad \eqref{C1}, \eqref{C3}\text{-}\eqref{C7}, \eqref{C9}, \notag
\end{align}
Note that the problem $\text{(P3.1)}$ is still a non-convex problem due to the non-convex  term $E_k^{fly}$ in objective function and non-convex constraint in \eqref{P3.1}. In the following, we adopt the SCA technique for solving the trajectory optimization problem. To this end, we define ${R_k}\left[ n \right] = {\log _2}\left( {1 + \frac{{{p_k}\left[ n \right]{h_k}\left[ n \right]}}{{{\sigma ^2} + \sum\limits_{k' \ne k} {{p_{k'}}\left[ n \right]{h_{k'}}\left[ n \right]} }}} \right)$, and $R_k[n]$ can be rewritten as
\begin{align}
{R_k}\left[ n \right] = {{\bar R}_k}\left[ n \right] - {\log _2}\left( {{\sigma ^2} + \sum\limits_{k' \ne k} {{p_{k'}}\left[ n \right]{h_{k'}}\left[ n \right]} } \right), \label{P.20}
\end{align}
where ${{\bar R}_k}\left[ n \right] = {\log _2}\left( {{\sigma ^2} + \sum\limits_{k \in \cal K} {{p_k}\left[ n \right]{h_k}\left[ n \right]} } \right)$.
Define $Q_k[n]=\|{\bf q}_k[n]-{\bf w}\|^2$, $k \in \cal K$, then the constraint \eqref{P3.1} can be rewritten as
\begin{align}
{{\bar R}_k}\left[ n \right] - {\log _2}\left( {{\sigma ^2} + \sum\limits_{k' \ne k} {\frac{{{p_{k'}}\left[ n \right]{\beta _0}}}{{{Q_{k'}}\left[ n \right] + {H^2}}}} } \right) \ge \frac{{{L_k}\left[ n \right]}}{{\delta B}}, \label{P3.2}
\end{align}
and
\begin{align}
{Q_k}[n] \le {\left\| {{{\bf{q}}_k}[n] - {\bf{w}}} \right\|^2}, \forall k, n\in \cal N. \label{P3.3}
\end{align}
While the term  ${\log _2}\left( {{\sigma ^2} + \sum\limits_{k' \ne k} {\frac{{{p_{k'}}\left[ n \right]{\beta _0}}}{{{Q_{k'}}\left[ n \right] + {H^2}}}} } \right)$ in \eqref{P3.2} is convex with respect to $Q_k[n]$, the term ${{\bar R}_k}\left[ n \right]$ is not convex with respect to ${\bf q}_k[n]$. In addition, it also  introduces a non-convex set in \eqref{P3.3}. To tackle the non-convex constraints \eqref{P3.2} and \eqref{P3.3}, we apply the SCA technique. For the term ${{\bar R}_k}\left[ n \right]$,  we  take the first-order Taylor expansion of ${{\bar R}_k}\left[ n \right]$ with respect to any given point ${\left\| {{{\bf{q}}_{l,k}}[n] - {\bf{w}}} \right\|^2}$ over $l \text{-} \rm th$ iteration,  we thus have the following inequality \eqref{P3.4} (see the inequation on top of the next page), where ${U_{l,k}}\left[ n \right] = \frac{{\frac{{{p_k}\left[ n \right]\beta {{\log }_2}e}}{{{{\left( {{{\left\| {{{\bf{q}}_{l,k}}[n] - {\bf{w}}} \right\|}^2} + {H^2}} \right)}^2}}}}}{{{\sigma ^2} + \sum\limits_{k \in \cal K} {\frac{{{p_k}\left[ n \right]\beta }}{{{{\left( {{{\left\| {{{\bf{q}}_{l,k}}[n] - {\bf{w}}} \right\|}^2} + {H^2}} \right)}^2}}}} }}$.
\newcounter{mytempeqncnt0}
\begin{figure*}
\normalsize
\setcounter{mytempeqncnt0}{\value{equation}}
\begin{align}
{{\bar R}_k}\left[ n \right] \ge {\log _2}\left( {{\sigma ^2} + \sum\limits_{k \in K} {\frac{{{p_k}\left[ n \right]{\beta _0}}}{{{{\left\| {{{\bf{q}}_{l,k}}[n] - {\bf{w}}} \right\|}^2} + {H^2}}}} } \right) - \sum\limits_{k = 1}^K {{U_{l,k}}\left[ n \right]\left( {{{\left\| {{{\bf{q}}_k}[n] - {\bf{w}}} \right\|}^2} - {{\left\| {{{\bf{q}}_{l,k}}[n] - {\bf{w}}} \right\|}^2} } \right)}  \overset{f} = \bar R_k^{lb}\left[ n \right] \label{P3.4}
\end{align}
\hrulefill 
\vspace*{4pt} 
\end{figure*}
Then, \eqref{P3.2} can be  replaced by its lower bound approximation result as
\begin{align}
&{{\bar R}_k^{lb}}\left[ n \right] - {\log _2}\left( {{\sigma ^2} + \sum\limits_{k' \ne k} {\frac{{{p_{k'}}\left[ n \right]{\beta _0}}}{{{Q_k}\left[ n \right] + {H^2}}}} } \right) \ge \frac{{{L_k}\left[ n \right]}}{{\delta B}}, \notag\\
& \qquad\qquad\qquad\qquad\qquad\qquad\qquad\qquad \forall k, n \in \cal N. \label{P3.5}
\end{align}
It is not difficult to verify that constraint \eqref{P3.5} is now convex. To tackle the  non-convex \eqref{P3.3}, a local convex approximation is still applied. Specifically, for   given local point ${\bf q}_{l,k}[n]$ over $l \text{-} \rm th$ iteration, we have
\begin{align}
{\left\| {{{\bf{q}}_k}[n] - {\bf{w}}} \right\|^2} \ge {\left\| {{{\bf{q}}_{l,k}}[n] - {\bf{w}}} \right\|^2} + 2{\left( {{{\bf{q}}_{l,k}}[n] - {\bf{w}}} \right)^T}\notag\\
\times \left( {{{\bf{q}}_k}[n] - {\bf{w}}} \right) \overset{g} = \varphi \left( {{{\bf{q}}_k}[n]} \right).
\end{align}
Define the new constraint
\begin{align}
{Q_k}[n] \le \varphi \left( {{{\bf{q}}_k}[n]} \right), \forall k, n \in \cal N. \label{P3.6}
\end{align}
For non-convex  term $E_k^{fly}$ in objective function of $\text{(P3.1)}$, it can still be replaced by a convex form by introducing the slack variable $\tau_n$, which can refer to Section III. As a result, for any given point $\left\{ {{{\bf{q}}_{l,k}}\left[ n \right]} \right\}$, define the following optimization problem
\begin{align}
&\text{(P3.2)}:\mathop {\min }\limits_{\scriptstyle\left\{ {A,B,\tau_n} \right\}\hfill\atop
\scriptstyle\left\{ {{\rho _k},Q_k[n]} \right\}\hfill} \sum\limits_{k = 1}^K {\bar E_k^{fly} + \sum\limits_{k = 1}^K {E_k^{comp} + \sum\limits_{k = 1}^K {\sum\limits_{n = 1}^{N - 1} { E_k^{com}\left[ n \right]} } } }  \notag \\
&{\rm{s.t.}} ~\eqref{C1}, \eqref{C3}\text{-}\eqref{C7}, \eqref{C9}, \eqref{NEWspeed}, \eqref{P3.5}, \eqref{P3.6}\notag
\end{align}
Problem $\text{(P3.2)}$ is now convex, which can be efficiently solved by standard convex optimization technique. Note that it can be readily verified that the objective value of $\text{(P3.2)}$  gives a upper  bound result to  that of problem $\text{(P3.1)}$.
\subsection{ Power Allocation Optimization with Fixed  UAV Trajectory}
In this section, we consider the second subproblem of problem $\text{(P3)}$ in which the transmission power of UAV is optimized while the UAV trajectory is fixed. The problem can be reformulated as
 \begin{align}
&\text{(P3.3)}:\mathop {\min }\limits_{\left\{ {A,{p_k}\left[ n \right],{\rho _k}} \right\}} \sum\limits_{k = 1}^K {E_k^{comp}}  + \sum\limits_{k = 1}^K {\sum\limits_{n = 1}^{N - 1} {E_k^{com}} \left[ n \right]}   \notag \\
&{\rm{s.t.}} ~{\log _2}\left( {1 + \frac{{{p_k}\left[ n \right]{h_k}\left[ n \right]}}{{{\sigma ^2} + \sum\limits_{k' \ne k} {{p_{{  k'}}}\left[ n \right]{h_{{ k'}}}\left[ n \right]} }}} \right) \ge \frac{{{L_k}\left[ n \right]}}{{\delta {B}}},\forall k,n \in \cal N,\label{P3.7}\\
&\qquad \eqref{C1}, \eqref{C3}\text{-}\eqref{C5}, \eqref{C7}\text{-}\eqref{C9}. \notag
\end{align}
Problem $\text{(P3.3)}$ is a non-convex optimization problem due to the non-convex constraint in \eqref{P3.7}.
Define $\tilde R_k[n]={\log _2}\left( {{\sigma ^2} + \sum\limits_{k' \ne k} {{p_{k'}}\left[ n \right]{h_{k'}}\left[ n \right]} } \right)$. To tackle the non-convex constraint  of \eqref{P3.7}, we apply the successive convex optimization technique to approximate $\tilde R_k[n]$ with a convex function in each iteration. Specifically, for any given local point $p_{l,k}[n]$ over $l \text{-} \rm th$ iteration, we have the inequality given in \eqref{P3.8} (see the inequation on top of the next page).
\newcounter{mytempeqncnt1}
\begin{figure*}
\normalsize
\setcounter{mytempeqncnt1}{\value{equation}}
\begin{align}
 \tilde R_k[n] \le {\log _2}\left( {{\sigma ^2} + \sum\limits_{k' \ne k} {{p_{l,k'}}\left[ n \right]{h_{k'}}\left[ n \right]} } \right) + \sum\limits_{k' \ne k} {\frac{{{h_{k'}}\left[ n \right]{{\log }_2}\left( e \right)}}{{{\sigma ^2} + \sum\limits_{k' \ne k} {{p_{l,k'}}\left[ n \right]{h_{k'}}\left[ n \right]} }}} \left( {{p_k}\left[ n \right] - {p_{l,k}}\left[ n \right]} \right) \overset {g}= \tilde R_k^{up}\left[ n \right] \label{P3.8}
\end{align}
\hrulefill 
\vspace*{4pt} 
\end{figure*}
As a result, for any given point $ {{{p}_{l,k}}\left[ n \right]} $,  problem $\text{(P3.3)}$ is approximated as
 \begin{align}
&\text{(P3.4)}:\mathop {\min }\limits_{\left\{ {A,{p_k}\left[ n \right],{\rho _k}} \right\}} \sum\limits_{k = 1}^K {E_k^{comp}}  + \sum\limits_{k = 1}^K {\sum\limits_{n = 1}^{N - 1} {E_k^{com}} \left[ n \right]}   \notag \\
&{\rm{s.t.}} ~\bar R_k[n]-\tilde R_k^{up}[n] \ge \frac{{{L_k}\left[ n \right]}}{{\delta {B}}},\forall k,n \in \cal N,\\
&\qquad \eqref{C1}, \eqref{C3}\text{-}\eqref{C5}, \eqref{C7}\text{-}\eqref{C9}. \notag
\end{align}
The problem $\text{(P3.4)}$  is a convex optimization problem, which can be efficiently solved by standard optimization technique. As a consequence, $\text{(P3.3)}$ can be approximated solved by successively updating the power allocation based on the optimal solution to $\text{(P3.4)}$ . It should be pointed out that the obtained solution by solving  problem $\text{(P3.4)}$  can be served as the upper bound of problem $\text{(P3.3)}$.

Based on the solutions to its two subproblems obtained by optimizing UAV trajectory and power allocation  via solving $\text {(P3.2)}$ and $\text {(P3.4)}$, we propose an iterative algorithm for problem  $\text {(P3)}$, which is summarized in Algorithm 2. The proof of convergence of Algorithm 2 can refer to Algorithm 1 provided in Section V.
\begin{algorithm}
\caption{ Joint optimization of UAV  trajectory and power allocation }
\label{alg1}
\begin{algorithmic}[1]
 \STATE  \textbf{Initialize} the UAV-TBS scheduling $p_k^{m}[n]$, iterative number $m=0$, and  a maximum iterative number $M_{max}$.
 \STATE  \textbf{Repeat}
\STATE  \qquad Solve  problem $\text {(P3.2)}$ for given  power allocation  \\
\qquad  $p_k^{m}[n]$, and denote the optimal UAV trajectory as \\
\qquad  $B^{m+1}$.
\STATE \qquad Solve  problem $\text {(P3.4)}$ for given  UAV trajectory $B^{m+1}$\\
\qquad   and denote the optimal UAV-TBS scheduling as \\
\qquad $p_k^{m+1}[n]$.
\STATE \qquad m=m+1.
\STATE  \textbf{Until}  a maximum iterative number has been reached or convergence.
\end{algorithmic}
\end{algorithm}
 \begin{figure}[!t]
\centerline{\includegraphics[width=3.2in]{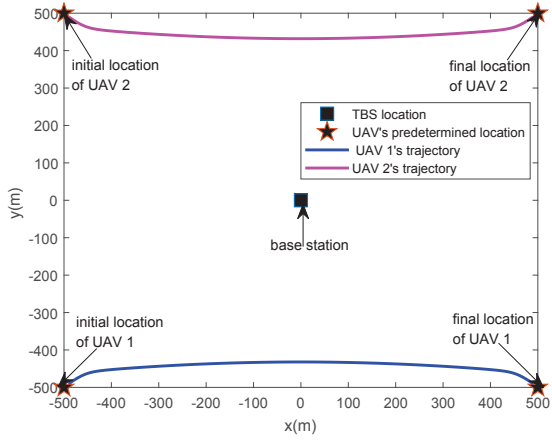}}
\caption{UAVs' trajectories  under  TDMA scheme with $\rm T=60\rm{s}$.} \label{fig2}
\end{figure}
 \begin{figure}[!t]
\centerline{\includegraphics[width=3.2in]{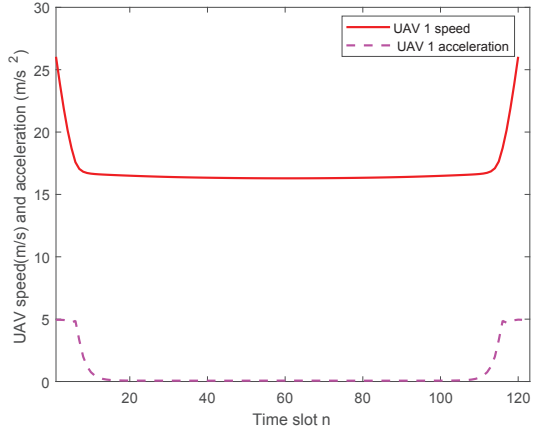}}
\caption{UAV 1 speed and acceleration  under TDMA scheme  with $\rm T=60\rm{s}$.} \label{fig3}
\end{figure}
 \begin{figure}[!t]
\centerline{\includegraphics[width=3.2in]{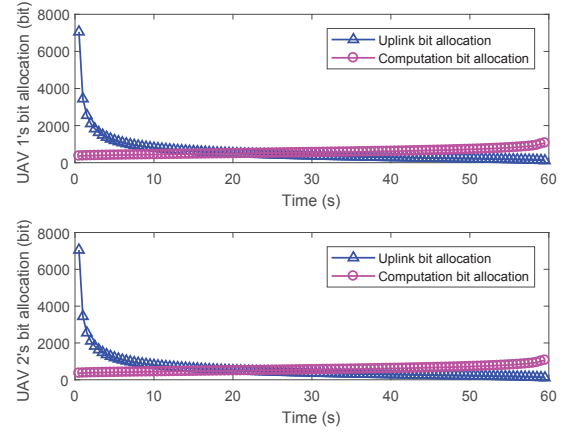}}
\caption{Optimized bit allocation for the UAV trajectory under TDMA scheme  with $\rm T=60\rm{s}$.} \label{fig4}
\end{figure}
 \begin{figure}[!t]
\centerline{\includegraphics[width=3.2in]{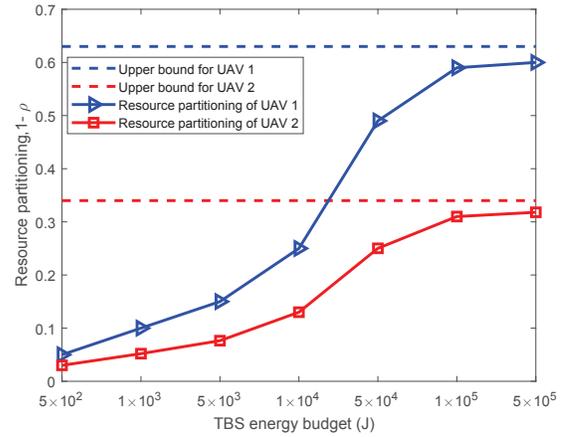}}
\caption{Optimized resource partitioning versus UAV energy budget under  TDMA scheme with $\rm T=60\rm{s}$.} \label{fig5}
\end{figure}
 \begin{figure}[!t]
\centerline{\includegraphics[width=3.2in]{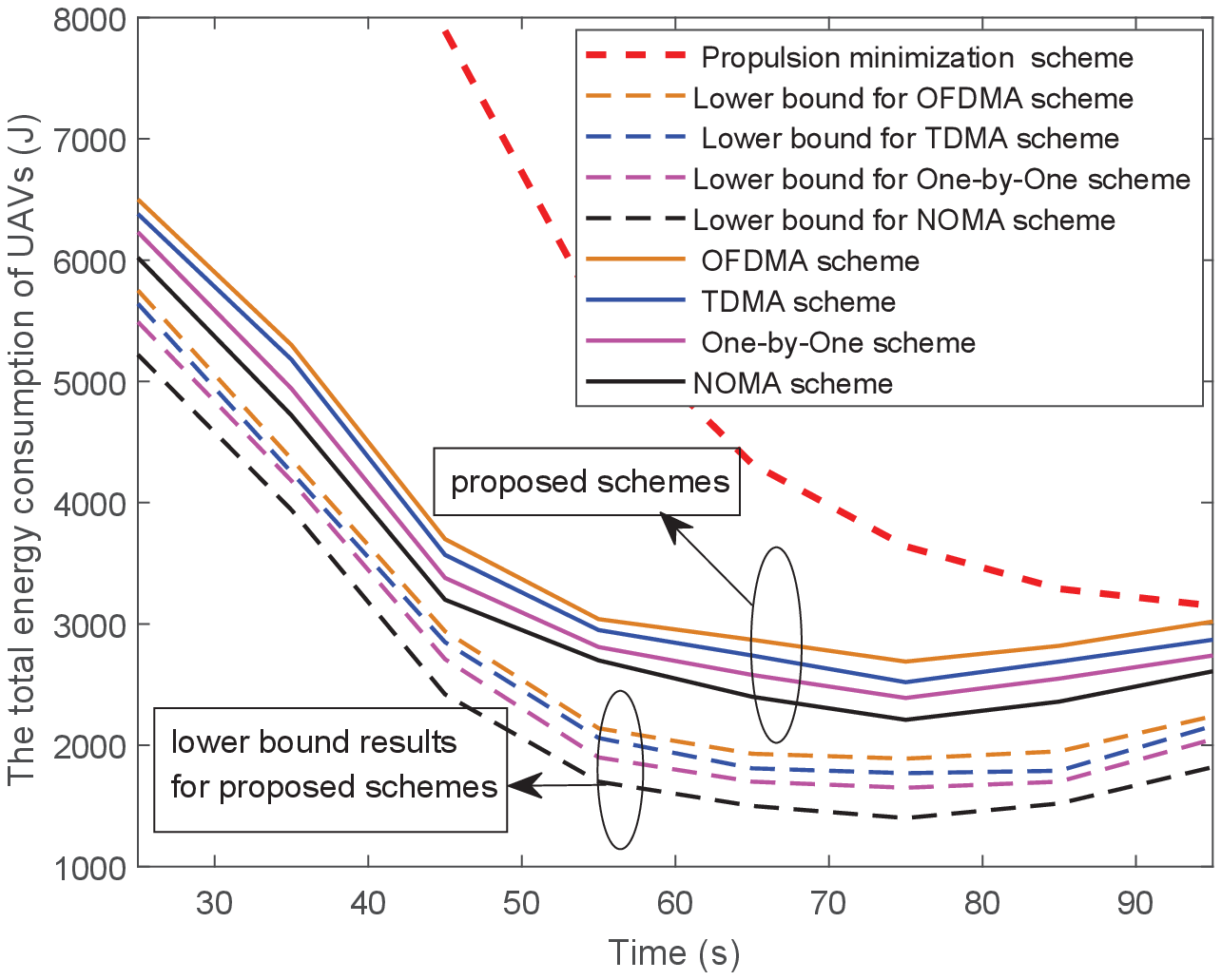}}
\caption{The total energy consumption of UAVs  under  TDMA scheme with $\rm T=60\rm{s}$ and $E_{total}=4\times 10^3 \rm{J}$.} \label{fig6}
\end{figure}
\section{NUMERICAL RESULTS}
In this section, we present the numerical results to illustrate the effectiveness of our proposed four schemes. In the simulations, the altitude of UAV is fixed at $H=80\rm{m}$ with the maximum transmission power $P_{max}=2\rm{W}$. The communication bandwidth is $B=1\rm{MHz}$ with the reference SNR ${\gamma _0} = 5\times10^3\rm{W}$. We assume that the maximum  energy budget of TBS  allocated for computing is $E_{total}=4\times10^3\rm{J}$ unless otherwise specified. The total number of bits for UAV 1 and UAV 2 are set to be $0.5\rm{Mbits}$ and $1\rm{Mbits}$, respectively. We assume that there have two UAVs, i.e., $K=2$, which can move to any direction subject to the maximum UAV speed $50\rm{m/s}$ and acceleration $5\rm{m/s^2}$. In addition, we assume that  $c1 = 0.002$, $c2=70.698 $ as in \cite{8329973} and \cite{Choi2014Energy}, and  the computation coefficient  is set to $G=10^{-11}$ as stated in \cite{sheng2015energy}. In addition, the time slot is assumed to be $\delta=0.5\rm{s}$. It is worth mentioning  that we plot the curves of UAV trajectory, bit allocation and resource partitioning for TDMA scheme, and the curves of UAV trajectory, bit allocation and resource partitioning for other schemes are not plotted. This is because the similar results can be obtained compared  with TDMA scheme.

In Fig.~\ref{fig2},  we plot the optimized UAVs' trajectories obtained by the TDMA  scheme with horizon time $\rm T=60\rm{s}$. We let UAV 1's initial location and final location respectively as  ${{\bf{q}}_{1,{\rm{I}}}} = {\left[ { - 500, - 500} \right]^T}$ and ${{\bf{q}}_{1,{\rm{F}}}} = {\left[ { 500, - 500} \right]^T}$ with corresponding UAV 1' initial velocity ${v_1}\left[ 0 \right] = {\left[ {20,20} \right]^T}(\rm{m/s})$ and final velocity ${v_1}\left[ N+1 \right] = {\left[ {20,-20} \right]^T}(\rm{m/s})$, and  UAV 2's initial location and final location respectively as  ${{\bf{q}}_{2,{\rm{I}}}} = {\left[ { - 500, 500} \right]^T}$ and ${{\bf{q}}_{2,{\rm{F}}}} = {\left[ { 500, 500} \right]^T}$ with corresponding UAV 2' initial velocity ${v_2}\left[ 0 \right] = {\left[ {20,-20} \right]^T}(\rm{m/s})$ and final velocity ${v_2}\left[ N+1 \right] = {\left[ {20,20} \right]^T}(\rm{m/s})$. In addition,  the horizon coordinate of TBS is located at ${\bf w}=(0,0)^T$. Three  prominent  insights can be achieved in  Fig.~\ref{fig2}. Firstly, the curves of UAVs' trajectories  for minimizing the total UAVs energy consumption   yield a straight flight mostly, which indicates that the energy-minimization strategy  is simply  straight flight manner with constant UAV speed. Secondly, for offloading bits from UAVs to TBS as more as possible, the UAVs tend to adjust itself location to shorten the distance between UAV and TBS. Thirdly, the propulsion-related energy consumption of UAVs is dominated compared with communication-related energy consumption, otherwise the UAVs prefer to move directly to TBS and hover above TBS for saving communication energy. For describing more detailed information about UAV's mobility, the curves of UAV 1 speed and acceleration are plotted in Fig.~\ref{fig3}. We can observe that UAV 1 firstly flies at a higher  speed, and then flies at a constant lower speed. In addition, the UAV 1 acceleration  almost equal to 0 from time slot $n=16$ to $n=108$, which results in a low energy consumption for flying. The results of speed and acceleration for UAV 2 have similar results as UAV 1, and is not discussed here for brevity.

Fig.~\ref{fig4} shows the achieved bit allocation of   uplink transmission phase and TBS local computation phase for UAV 1 and UAV 2. From the first sub-figure of Fig.~\ref{fig4} for UAV 1, we can see that a large number of bits are  allocated in the uplink transmission for UAV 1 at begin, and then the   number of bits are allocated in  uplink transmission for UAV 1 decrease. This conclusion is contrary to the results stated in \cite{Jeong2018Mobile}, which shows that  as the UAV is closer to mobile user, a larger number of bits should be allocated.
This is attributed to the fact that the  work  \cite{Jeong2018Mobile} focuses on  minimizing  the mobile users' communication energy, however, our work  pays attention to minimizing the total energy consumption of UAVs, including communication-related energy consumption, local computation energy consumption and propulsion-related energy consumption. While as the distance between UAV and TBS become shorter, the  larger number of bits allocated for TBS will reduce the communication-related energy consumption of UAVs, the energy used for flying and computing will increase. For TBS local computation phase, the TBS prefers to process the equal number of bits in each time slot for saving TBS' computation energy, which has same conclusion as in \cite{Jeong2018Mobile}. Moveover, we can see that from the second  sub-figure of Fig.~\ref{fig4} for UAV 2, the number of bits allocated for uplink transmission and computation is almost same as in the sub-figure of Fig.~\ref{fig4} for UAV 1. The reason can be explained as  the energy budget for computation is limited, thus it only can process a limited number of bits received from UAVs.

Fig.~\ref{fig5} shows the impacts of TBS energy budget on UAV' resource partitioning strategy. Two dash lines, i.e., blue dash line and red dash line, represent  the maximum resource partitioning values for  UAV 1 and UAV 2, respectively.  The maximum resource partitioning values  are derived from the special case for TDMA which discussed in Fig.~\ref{fig6}. In Fig.~\ref{fig5},  we can observe that the optimized resource partitioning, $1-\rho_k, \forall k$, is monotonically increasing with TBS energy budget, which means that the more number of bits of UAVs can be  offloaded to TBS for computation as TBS energy budget increases. This can be explained  as the more TBS energy budget, the more number of bits can be processed. Furthermore, it can be observed that the resource partitioning of UAV 1 is higher than UAV 2. This is because from Fig.~\ref{fig4}, it  shows that the number of bits of two UAVs allocated for TBS  is almost same, and also the total number of bits of UAV 1 is  smaller than the total number of bits of UAV 2, thus the results can be directly obtained.

At last, we compare our proposed four schemes with comparison benchmarks  in terms of minimizing the total energy consumption of UAVs as shown in Fig.~\ref{fig6}. For  propulsion minimization scheme, the UAV's trajectory is firstly optimized for minimizing the UAV's propulsion-related energy, then with the obtained UAV trajectory, we optimize bit allocation and resource partitioning of UAV for jointly  minimizing  communication-related energy and local computation energy. We also  consider four special schemes as comparison benchmarks  which the TBS allocated energy budget  for computation is sufficient large and the \emph{bit-casuality } constraint (namely we  assume that the TBS can completely process the bits  received from UAVs within one time slot) is ignored. Obviously, these four special benchmarks can be served as lower bound results for the proposed four schemes. In Fig.~\ref{fig6}, it  can be observed that the  adopted  propulsion minimization scheme consumes  more energy    than the other schemes. We can conclude that   the propulsion-related UAV energy consumption is not the only  prominent factor for the total UAV energy consumption, it also indicates that the bits used for  computation consume a large amount of energy.
It can also be   see that the NOMA scheme has better  performance than the orthogonal schemes (TDMA, OFDMA, One-by-One), the larger gain for NOMA scheme can be attributed to  the property of sharing of entire  time and bandwidth resource simultaneously among UAVs. However, the successive interference cancellation (SIC) technique used for NOMA  may bring an additional interference and implementation complexity  compared with orthogonal schemes \cite{Ding2014On},\cite{zhu2017optimal}. Furthermore,  the gap among proposed four  schemes is not large, thus, there exists a tradeoff between implementation complexity and performance gain. In addition, we can see that the  four solid curves  plotted  in  Fig.~\ref{fig6} firstly decrease with horizon time $\rm T$ from  $\rm {T=25s} $  to  $\rm {T=75s} $, and increase with horizon time $\rm T$ from  $\rm {T=75s} $  to  $\rm {T=95s} $. This can be explained since the  horizon time $\rm T$  is small in the first phase, the energy consumption used for  computation  is larger than the UAV used for flying; In the second phase, as  horizon time $\rm T$ becomes larger,  the energy consumption used for flying   is larger than the UAV used for  computation; In  other words,  the energy of UAV used for flying increases with  horizon time $\rm T$ and the energy of UAV used for computation decreases with time $\rm T$. Therefore, there exists a tradeoff between propulsion energy consumption of UAVs  and local computation energy consumption of UAVs, and the  optimal time $\rm T$ for minimizing the total energy consumption of UAVs is left in our future work.

\section{Conclusion}
In this paper, we propose four types of access schemes in the uplink transmission for mobile edge computing system.   We formulate the problem as a energy minimization problem while ensuring the large  number of bits of UAVs are completely computed in a given horizon time. For solving the  non-convex optimization problem, a sub-optimal result is  achieved by  using successive convex approximation  technique. The numerical results show that  there exists a tradeoff between propulsion energy consumption of UAVs  and computation energy consumption of UAVs. In addition, it also shows that our proposed four  schemes save a large amount of energy  compared with the propulsion minimization scheme.

%
%
%
%
%

\ifCLASSOPTIONcaptionsoff
  \newpage
\fi



%
%
%
\bibliographystyle{IEEEtran}
\bibliography{IEEEACCESS}

%

%
%
%




\end{document}